\documentstyle[aasms4]{article} 
\begin{document}
\title{Looking Down the Light Cone: Can Deep Redshift Surveys Alone 
Measure the Power Spectrum?}

\author{Andrew A.\ de Laix and Glenn D. Starkman}

\affil{Physics Department, Case Western Reserve University, Cleveland,
OH 44106-7079}

\authoremail{aad4@po.cwru.edu}

\begin{abstract} 
We analyze the window functions for the spherical harmonic mode
estimators of all--sky, volume limited surveys considering
evolutionary effects along the past light--cone which include the
deviation of the distance scale from a linear relationship with
redshift, linear peculiar velocity corrections, and linear evolution
of the density perturbations.  The spherical harmonic basis functions
are considered because they correspond most closely to the symmetries
of typical survey geometries and of the light--cone effects we
consider.  Our results show substantial broadening of the windows over
that expected by ignoring light--cone effects, indicating the
difficulty of measuring the power spectrum independently from
cosmology.  We suggest that because of light--cone effects, deep
redshift surveys should either be analyzed in conjunction with CMBR
data which determines the cosmological parameters, or by using a
Bayesian likelihood scheme in which varying cosmological parameters
and a simple parameterization of the primordial power spectrum are
assumed as the priors, so that observed data can be mapped from
redshift to real space.  The derived power spectrum can then be
compared to underlying models of fluctuation generation and growth in
structure formation to evaluate both these models and the cosmological
priors.

\end{abstract}

\section*{}
The ability of a number of current and proposed galaxy surveys, {\it
e.g}.\ The Sloan Digital Sky Survey (hereafter SDSS, \cite{gunn}) and
the two-degree-field survey (\cite{2df}), to probe deep redshifts
offers exciting opportunities to observe structure on the largest
scales, yielding results which will be invaluable for cosmology. As is
often the case, however, the data collected is not directly compatible
with that which theorists calculate. To see this, consider a typical
theoretical calculation of large scale structure.  Structure is
presumed to have evolved from the gravitational collapse of small
initial fluctuations, and on scales larger than approximately 10 Mpc,
this collapse is well described by linear perturbation theory.
Usually, theorists express the outcomes of their structure-formation
models in terms of the power spectrum, the Fourier transforms of the
two-point correlation function given in conformal spatial coordinates,
evaluated on a constant time slice.  That is, they measure the power
spectrum of the structure for a snapshot of the universe taken at a
fixed time.  A galaxy survey, however, detects structure on a
light--cone time slice, and measures redshifts, not conformal
distances.  For lengths much shorter than the Hubble length, the
distinction is unimportant since to first order distance is
proportional to redshift, and structure does not evolve much over the
light-crossing time of the survey.  But, the newest surveys, such as
the SDSS, will penetrate to as much as 50\% of the Hubble distance.
Either assuming that proper distance is linearly proportional to redshift
over these distance scales, or  ignoring the evolution of structure
over the corresponding time scales, leads to significant errors in the
estimation of the power spectrum.  In this letter, we examine these
effects and determine their significance for estimating the power
spectrum from a deep galaxy survey.

All surveys are limited in the volume of space which they can cover,
and so are less than ideal observations.  Furthermore, shot noise
arises from the finite sampling inherent in any galaxy
survey. Conversely, theoretical predictions for power spectra are
determined for the whole universe continuously sampled and are
therefore expressed in terms of a continuous power spectrum (unless
the universe is spatially compact, but no current evidence suggests
that it is).  In sampling a finite volume, one probes instead a
discrete set of eigenmodes and the amplitudes of the observed modes
will be a convolution of the continuum modes with a window function.
Ideally, this window function is narrow so that the observed mode
amplitudes correspond closely to the their continuum counterparts, and
this can be achieved through the proper weighting of the observational
data.

For Fourier decomposition, \cite{tegmark} has developed a method of
analyzing redshift surveys guaranteed for a specified survey geometry
to minimize the width of the window function to the shot noise limit.
Fourier decomposition, however, is not well suited to the analysis of
most redshift surveys because survey geometries are usually defined by
slices of a sphere, and because redshift effects depend only on the
distance from the observer.  A better basis is one which takes
advantage of this symmetry, {\it i.e}.\ the spherical harmonic basis
(\cite{heavens}).  For standard homogeneous, isotropic models, the power
spectrum of the spherical harmonic decomposition is equivalent to that
of the Fourier decomposition --- each depends only on the magnitude of
the mode considered, so theoretical calculations are equally useful
for each basis.  Our analysis therefore addresses the determination or
estimation of the power spectrum in the spherical harmonic basis.

Given a survey geometry, we would like to find a weighting function
$\psi$ which will produce the ``optimal'' estimate for the amplitude
of a given $\ell_0, m_0,$ and $k_0$ harmonic of the continuum
fluctuation spectrum.  By optimal we mean that $\psi$ should
simultaneously minimize the width of the window function in $\ell, m,$
and $k$ space and minimize the shot noise signal due to finite
sampling---under the constraint that $\psi$ is non--trivial.
Following \cite{tegmark}, we define the penalty function
\begin{equation}
\label{penalty}
	|\ell-\ell_0|^2 + |m - m_0|^2 + {|k^2 - k_0^2|^2 \over
	\gamma^2},	
\end{equation}
which forces our window toward the $\ell_0, m_0,$ and $k_0$ we are
trying to observe.  The shot noise is proportional to $n(z)^{-1}$,
where $n(z)$ is the average number of galaxies observed in a unit
volume at redshift $z$.  To find the optimal $\psi$ we solve the eigen
value problem
\begin{equation}
\label{eigen_prob}
	\left( |\hat{\ell}-\ell_0|^2 + |\hat{m} - m_0|^2 
	+ {|\hat{k}^2 - k_0^2|^2 \over
	\gamma^2} + {1 \over n(z) }\right) \psi = E \psi,
\end{equation}
where the ground state represents the optimal weighting function as it
minimizes $E$.
We normalize it, for convenience, by requiring 
\begin{equation}
\label{norm}
\int d^3z \psi^2(z) = 1.
\end{equation}
Note that $\psi$ can always be chosen to be a real valued function
(which we shall assume it is), simplifying the numerics.  Also note
that for the radial portion of the penalty function we choose
$|\hat{k}^2 - k_0^2|^2$ instead of $|\hat{k} - k_0|^2$, which one
might naively select.  We do so because there is no simple way to
express $\hat{k}$ in coordinate space; the operator $\hat{k}^2$ is
simply negative the Laplacian in physical space. The operators
$\hat{\ell}$ and $\hat{m}$ return the magnitude of the total angular
momentum and the $z$ component respectively.  Since the angular
portion is separable, we can re-express $\psi$ as a product $\psi_{\mu
\nu k}(z) Y_{\mu \nu}(\Omega_z)$ where $\mu$ and $\nu$ are the 
eigenvalues of the generalized spherical harmonics which satisfy the
boundary conditions given by the survey geometry.  In the limit of an
all sky survey, $\mu = \ell_0$ and $\nu = m_0$ which are integers,
although this not generally so.  The parameter $\gamma$ controls how
we wish to weight shot noise to the window width.  By increasing
$\gamma$, we can reduce shot noise by accepting a larger variation in
$k$ , or by reducing $\gamma$ we can get a narrower $k$ variation, but
at the expense of increased shot noise.  For the special case of a
volume limited survey in which $n(z)$ is a constant, the shot noise
can be absorbed into the right hand side of the equation and the
choice of $\gamma$ is irrelevant; there is only one optimal $\psi$.  
As we have already pointed out, the angular part of
the solution is given by the generalized spherical harmonics. The
solution to the radial portion is then just a spherical Bessel
function $j_\mu(k_\mu z)$, where $j_\mu(k_\mu z_{max}) = 0 $ at the
outer boundary of the volume $z_{max}$, and $k_{\mu}$ is the zero
closest to $k$, the wavenumber of the mode whose amplitude we wish to
estimate.  It is this simplified problem which we will consider in
this letter since the numerics are easier, and we expect that any
light--cone effects which we observe in a more general survey of
similar depth will be at least if not more severe.

Given the optimal weighting function, our estimator for the
fluctuation amplitude of the $k, ~\ell$, $m$ mode is
\begin{equation}
\label{estimator}
	\hat{a}_{l m}(k) = \int d^3z {n(\vec{z})- \bar{n}(\vec{z})\over
	\bar{n}(\vec{z})} \psi(\vec{z}),
\end{equation}
where $\bar{n}(\vec{z}) = \langle n(\vec{z}) \rangle$ is the ensemble
average of the number density of galaxies at redshift $z$.  A more
interesting quantity is the square of this estimator, which averaged
over the ensemble of realizations should be related to the
continuum power spectrum.  Using the analysis of \cite{peebles}, one
can straight forwardly show that in the ensemble average
\begin{equation}
\label{est_sq}
	\left\langle  \hat{a}_{l m}(k) \hat{a}_{l m}(k) \right\rangle
	= \int d^3z {\psi^2(\vec{z}) \over 
	\bar{n}(\vec{z})} +
	\int d^3z d^3z' \xi(\vec{z},\vec{z}') \psi(\vec{z})
	\psi(\vec{z}'),
\end{equation}
where $\xi$ is the correlation function in redshift space.  From the
above, we can infer an estimator for the power in a particular mode
\begin{equation}
\label{P_lm_est}
	\hat{P}_{\ell m}(k) \equiv  
	\hat{a}_{l m}(k) \hat{a}_{l m}(k) 
	- \int d^3z {\psi^2(\vec{z}) \over 
	\bar{n}(\vec{z})},
\end{equation} 
where we subtract out the expected shot noise contribution.  The light
cone projection of the correlation function for flat space
(hyperbolic, or ``open'' models will be covered in a more detailed
paper \cite{delaix1}) has been calculated by \cite{nakamura}. In terms
of a spherical harmonic basis, $\xi$ may be expressed as
\begin{eqnarray}
\label{correlation}
	\xi(\vec{z},\vec{z}') &=& \sum_{\ell m} \int {2 k^2 dk \over
	\pi} \bar{D}(z)\bar{D}(z') Y_{\ell m}(\Omega_z) Y_{\ell
	m}(\Omega_z') 
	\\ \nonumber&\times&
	\Bigg\{ j_\ell(kr)-{\beta(z) \over k^2} \Bigg[ {H(z) \over
	1+z} \left({k\ell \over 2\ell+1} j_{\ell-1}(kr)- 
	{k(\ell+1) \over 2\ell+1}j_{\ell+1}(kr) \right)
	\\ \nonumber &-& 
	k^2\left( A_\ell j_{\ell-2}(kr) -B_\ell  j_{\ell}(kr)
	+ C_\ell  j_{\ell-2}(kr) \right) \Bigg] \Bigg\}
	\Bigg\{ z \rightarrow z' \Bigg \},
\end{eqnarray}
with
\begin{eqnarray*}
	A_\ell &=& {\ell (\ell-1) \over (2\ell+1)(2\ell-1)}\\
	B_\ell &=& {\ell^2 \over (2\ell+1)(2\ell-1)} - 
	{(\ell+1)^2 \over (2\ell+1)(2\ell+3)}\\
	C_\ell &=& {(\ell+1)(\ell+2) \over (2\ell+1)(2\ell+3)}.
\end{eqnarray*}
The proper radial distance $r$ is
\begin{equation}
\label{radial}
r = \int_0^z {dz \over H(z)},
\end{equation}
where we introduce the redshift--dependent
Hubble parameter for flat space 
\begin{equation}
\label{hubble}
	H(z) =  H_0 \sqrt{\Omega_0 (1+z)^3 + \lambda_0} .
\end{equation}
The parameters $\Omega_0$ and $\lambda_0$ represent respectively the
present day matter density and present day cosmological constant, in
units of the critical density, and $\Omega_0 + \lambda_0 = 1$ for flat
space.  We also use the linear growth factor
\begin{equation}
\label{growth}
D(z) = {5 \Omega_0 H_0^2 \over 2} H(z) \int_{z}^{\infty} {1+z' \over
H(z')^3} d z',
\end{equation}
along with the linearly evolving bias parameter which, following
\cite{fry}, can be written
\begin{equation}
\label{bias}
b(z) = 1 + {D(0) \over D(z)}(b_0 - 1),
\end{equation}
to define an overall linear growth factor for the galaxy perturbations
relative to the underlying density perturbations:
\begin{equation}
	\bar{D}(z) = {D(z)b(z) \over D(0)b(0)}. 
\end{equation}
Finally, we use linear velocity distortion parameter $\beta$ given by
\begin{equation}
\label{beta}
\beta(z) = -{d \ln D(z) \over d \ln (1+z)}{1 \over b(z)}.
\end{equation}
The reader is reminded that eq.\ (\ref{correlation}) is only valid
when $\Omega_0 + \lambda_0 = 1$, otherwise an expression with the
generalized spherical Bessel functions derived from a hyperbolic
Laplacian are required.

Using our expression for the correlation function, we can show
that the ensemble average of the estimator in eq.\ (\ref{P_lm_est}) 
is equal to 
\begin{equation}
\label{power_est}
	\langle \hat{P}_{\ell m}(k) \rangle =  \int k'^2dk' P(k')
	{2 \over \pi}\sum_{\ell' m'} I^2_{\ell' m'}(k'),
\end{equation}
with 
\begin{eqnarray}
\label{I}
	I_{\ell m}(k) &\equiv& \int d^3z \psi(\vec{z}) \bar{D}(z)  
	Y_{\ell m}(\Omega_z)\\ \nonumber
	&\times&
	\Bigg\{ j_\ell(kr)-{\beta(z) \over k^2} \Bigg[ {H(z) \over
	1+z} \left({k\ell \over 2\ell+1} j_{\ell-1}(kr)- 
	{k(\ell+1) \over 2\ell+1}j_{\ell+1}(kr) \right)
	\\ \nonumber &-& 
	k^2\left(A_\ell j_{\ell-2}(kr) -B_\ell  j_{\ell}(kr)
	+ C_\ell  j_{\ell-2}(kr) \right) \Bigg] \Bigg\}	
\end{eqnarray}
In the special case of an all sky survey, where $\psi(z) \propto
Y_{\ell_om_o}(\Omega_z)$, one can perform the angular integral
$d\Omega_z$ trivially.  The estimator for the power in a particular
mode is then reduced to the convolution of the true continuum power
spectrum with a window function
\begin{equation}
\label{window}
W_{\ell k} =  k^2 I_{\ell}^2(k), 
\end{equation}
which is independent of $m$.  It is these windows which indicate how
well a power spectrum derived from a limited redshift survey will
correspond to the underlying continuum power spectrum.

To demonstrate how light--cone effects will influence the
observational window functions defined above, we compare the windows
which arise when all redshift effects are included to those obtained
ignoring light-cone effects by taking the distance $r$ proportional to
the redshift ($r = H_0^{-1} z$) and holding the other redshift
dependent parameters --- $\beta$, $H$, and $\bar{D}$ -- fixed at their
$z=0$ values.  Often one sees results calculated in the latter naive
limit.  For simplicity we consider only volume-limited all--sky
surveys, investigating two survey depths: $z_{max} = 0.25$ and
$z_{max}=0.5$, which correspond to the approximate depths of the SDSS
magnitude-limited galaxy survey and the SDSS bright red galaxy survey.
We take unbiased ($b_0 = 1$), $\Omega_0 = 1$ cosmologies, and consider
two values of $k$, $0.033$ and .33 $h$ Mpc$^{-1}$.

In figure \ref{fig1}, we show a plot of the naive window functions for
$k = 0.033$ $h$ Mpc$^{-1}$ with a range of $\ell$'s in a $z <
0.25$ volume--limited survey, ignoring the light--cone effects, {\it
i.e}.\ this shows the optimal results one can hope to achieve for a
volume-limited survey of the specified depth.  In figure \ref{fig2},
we show the same results for a realistic survey, including light--cone
effects.  In figures \ref{fig3} and \ref{fig4}, we consider the same
survey geometry, but calculate the window functions for $k = 0.33~h$
Mpc$^{-1}$.  The former plot shown the naive window function while the
latter shows the realistic result.  Next we examine the deeper survey
geometry which extends to a depth of $z = 0.5$.  Again we consider
first the larger scale, $k = 0.033~h$ Mpc$^{-1}$, in the naive and
realistic limit in figures \ref{fig5} and \ref{fig6} respectively.
Finally, we examine the smaller scale in figures \ref{fig7} and
\ref{fig8}.

The results shown in the figures described above reveal the importance
of considering light--cone effects when estimating power from galaxy
surveys, with several effects meriting closer attention.  The most
striking is the broadening of the realistic window functions compared
to their idealized counterparts.  This is caused predominantly by the
deviation from the linear distance--redshift relation.  At higher
redshifts, the larger Hubble expansion rate causes the galaxies to
appear more spread out than they do locally.  Thus a wave that probes
a given scale locally, probes a smaller scale (larger $k$) at high
redshift.  A mode in conformal space will then couple to a range of
scales when mapped into redshift space, which leads to a broadening in
the window.  It also causes a readily observed shift in the peak to
larger scales. We can estimate the broadening quite easily if we
consider the effective frequency at which a conformal mode oscillates
in redshift space.  For large values of the argument, a spherical
Bessel function can be approximated $j_\ell(kr) \approx
\sin(kr+\ell\pi/2)/kr$.  In a small region $\Delta z$ located at $z$,
we can rewrite this as $\sin(kr(z)+ k\partial r/ \partial z\Delta z
+\ell\pi/2)/kr(z)$, which has an effective frequency of oscillation of
$ k\partial r/ \partial z$.  Recall from eq. (\ref{radial}) that
$\partial r/ \partial z = H^{-1}(z)$.  This tells us that the
effective frequency of the conformal modes which describe the density
fluctuations will drift from an initial value of $k$ to $
H^{-1}(z_{max})k$ in redshift space, so the inverse of this quantity
should be a good estimate of the actual window width.  For our
examples of $z_{max} = 0.25$ and $0.5$, we see that the window widths
should be about 1.4$k$ and 1.8$k$ respectively, which are in good
agreement with the results presented in the figures.
 A final effect, which is less visible, is an overall
decrease in the amplitude of the window function.  Since the power
spectrum decreases with growing $z$, the overall power measured in a
given mode will be less than its local value.

These results are important for analyzing observations, because they
indicate that it is impossible to make a clean measurement of the
power spectrum, independent of cosmology, from a deep redshift survey.
The light--cone effects result in significant broadening and shifting
of the window functions when compared to their idealized limit, in a
way that depends sensitively on the cosmology.  One must deconvolve
these effects to compare observations to theory, but this requires
knowledge of the underlying cosmological model (perhaps soon to be
provided by cosmic microwave background observations).  An alternative
approach might be to do a Bayesian analysis assuming different
cosmologies as priors.  One maps the observed data into conformal
coordinate space, utilizing the selected cosmology, calculates a power
spectrum estimate, and compares the estimate to theoretical
predictions based on simple parameterizations of the underlying power
spectrum from theories like inflation or defects.  For the correct
cosmological model, the windows will have minimum width, while the
incorrect models will have broadened windows.  It is possible that
these effects can be used to learn about the underlying cosmology, and
even to break the degeneracy in the power spectrum for models in
which the shape parameter, $\Omega h$, is constant (we are currently
investigating this problem).

\section*{Acknowledgments}
\noindent G.D.S. was supported by an NSF career grant.

\newpage
\section*{Figure Captions}

\figcaption{\label{fig1}The naive window function (no light--cone
effects) for an all--sky survey of depth $z=.25$ used to optimally
estimate the power in the $.033~h$ Mpc$^{-1}$ mode.  The results for
number of different $\ell$ modes are shown.}

\figcaption{\label{fig2}The true, light--cone corrected window
functions for an all--sky survey of depth $z=.25$ used to optimally
estimate the power in the $.033~h$ Mpc$^{-1}$ mode.  The results for
number of different $\ell$ modes are shown.}

\figcaption{\label{fig3}The naive window function (no light--cone
effects) for an all--sky survey of depth $z=.25$ used to optimally
estimate the power in the $.33~h$ Mpc$^{-1}$ mode.  The results for
number of different $\ell$ modes are shown.}

\figcaption{\label{fig4}The true, light--cone corrected window
functions for an all--sky survey of depth $z=.25$ used to optimally
estimate the power in the $.33~h$ Mpc$^{-1}$ mode.  The results for
number of different $\ell$ modes are shown.}

\figcaption{\label{fig5}The naive window function (no light--cone
effects) for an all--sky survey of depth $z=.5$ used to optimally
estimate the power in the $.033~h$ Mpc$^{-1}$ mode.  The results for
number of different $\ell$ modes are shown.}

\figcaption{\label{fig6}The true, light--cone corrected window
functions for an all--sky survey of depth $z=.5$ used to optimally
estimate the power in the $.033~h$ Mpc$^{-1}$ mode.  The results for
number of different $\ell$ modes are shown.}

\figcaption{\label{fig7}The naive window function (no light--cone
effects) for an all--sky survey of depth $z=.5$ used to optimally
estimate the power in the $.33~h$ Mpc$^{-1}$ mode.  The results for
number of different $\ell$ modes are shown.}

\figcaption{\label{fig8}The true, light--cone corrected window
functions for an all--sky survey of depth $z=.5$ used to optimally
estimate the power in the $.33~h$ Mpc$^{-1}$ mode.  The results for
number of different $\ell$ modes are shown.}


\begin{thebibliography}{}
\bibitem[]{surveys}

\bibitem[de Laix \& Starkman (1998)]{delaix1} de Laix, A. A. and
Starkman, G. D., in preparation

\bibitem[Fry (1996)]{fry} Fry, J.\ 1996, ApJ, 461, L5

\bibitem[Gunn \& Weinberg (1995), Strauss (1997)]{gunn} Gunn, J. E.,
\& Weinberg, D. H., in ``Wide Field Spectroscopy and the Distant
Universe'', ed.\ S. J. Maddox \& A. Arag\'{o}n--Salamanca (Singapore:
World Scientific); Strauss, M. A. 1997, preprint astro-ph/9610032 

\bibitem[Heavens and Talor (1995)]{heavens} Heavens, A. F. and Taylor, 
A. N. 1995, MNRAS, 278, 73

\bibitem[Nakamura, Matsubara, \& Suto (1997)]{nakamura}
 Nakamura, T. T., Matsubara, T., and Suto, Y. 1997, preprint,
 astro--ph/9706034

\bibitem[Peebles (1980)]{peebles} Peebles, P.\ J.\ E.\ 1980, The Large
Scale Structure on the Universe (Princeton: Princeton University
Press)

\bibitem[Taylor \& Gray (1994)]{2df} Taylor. K., Gray, P. M. 1994,
Proc. SPIE 2198, 136, ``Instrumentation in Astronomy VIII'', ed.\
Crawford, D. L. \& Craine, E. R.

\bibitem[Tegmark (1995)]{tegmark} Tegmark, M. 1995, ApJ, 455, 429

\end{thebibliography}
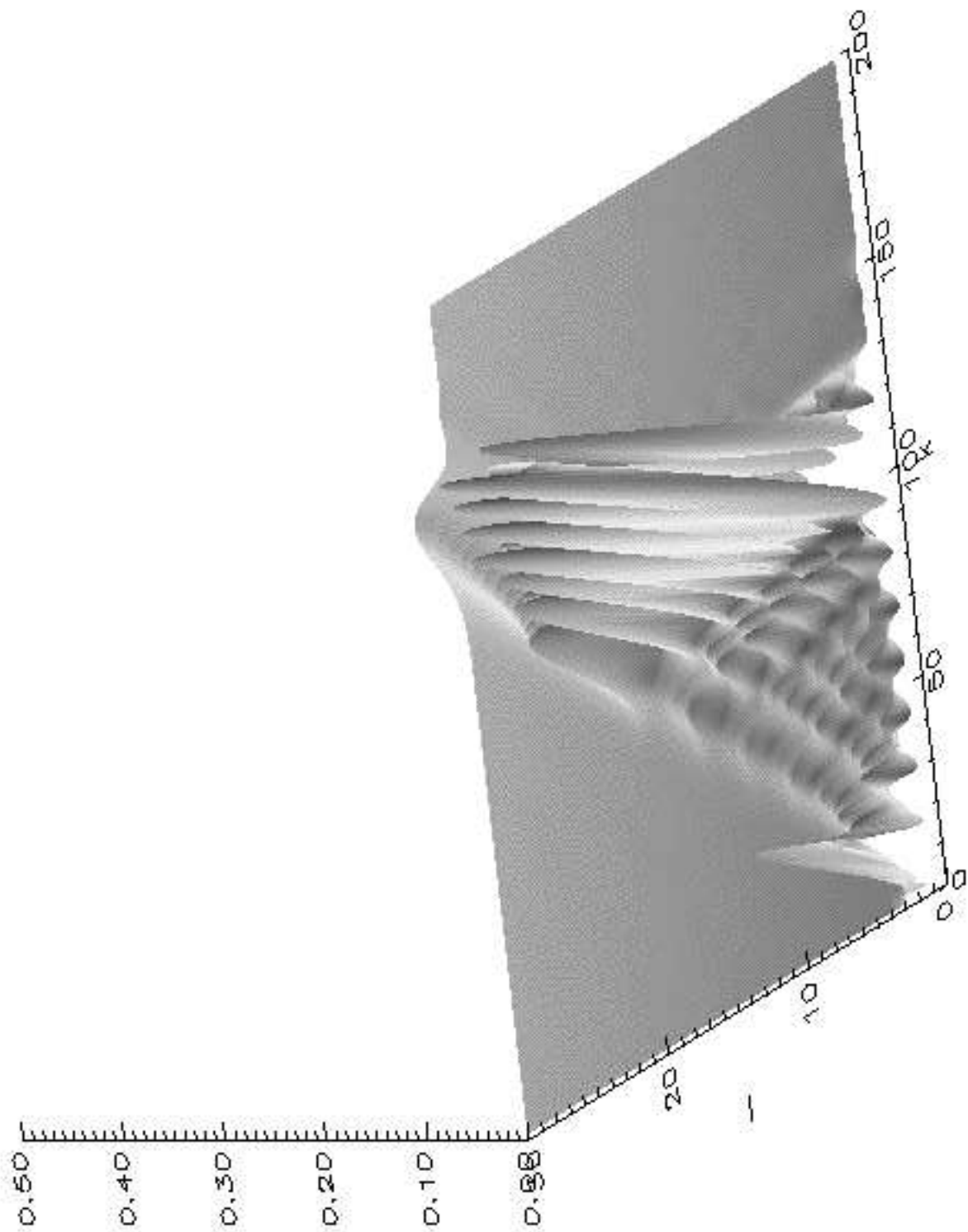
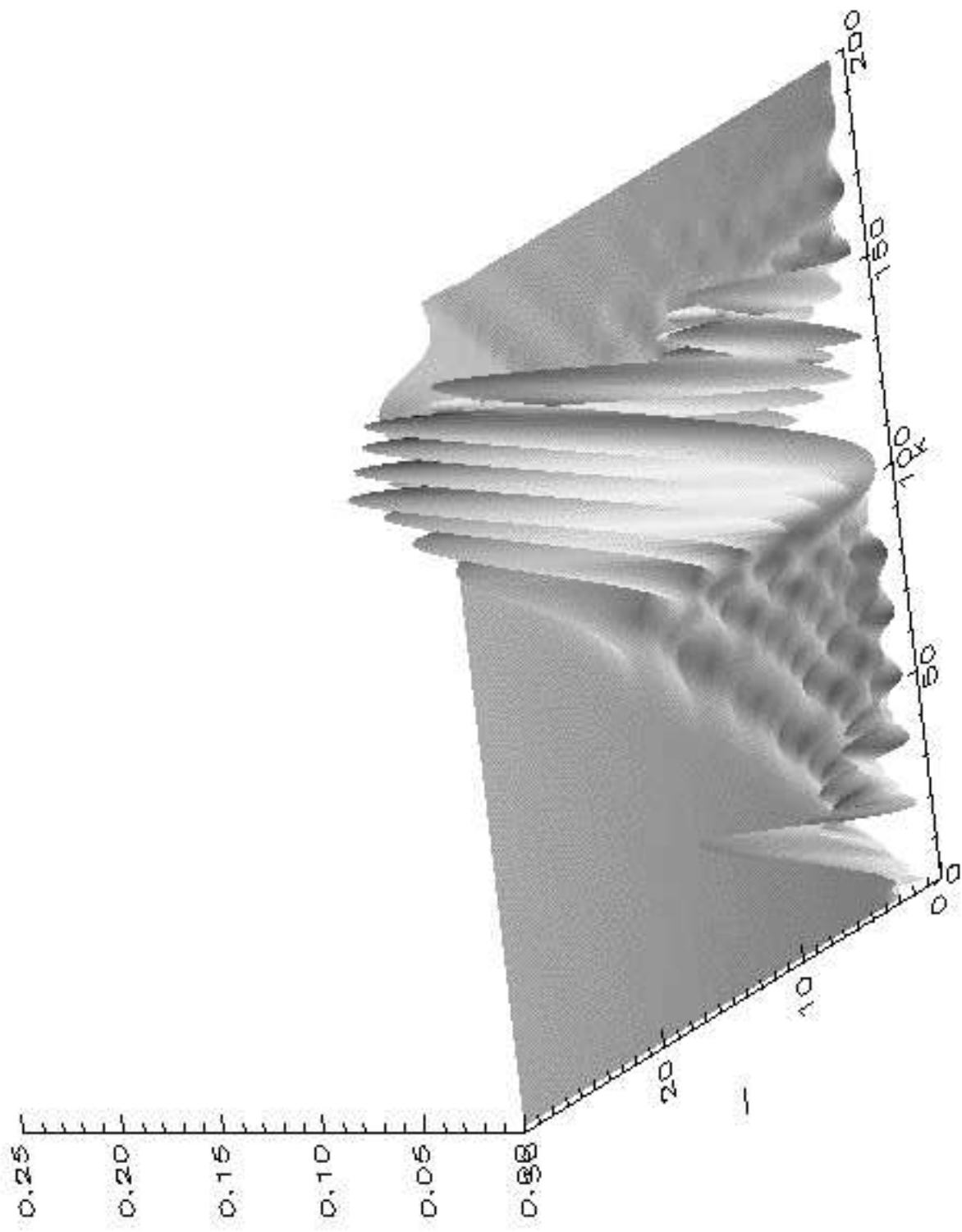
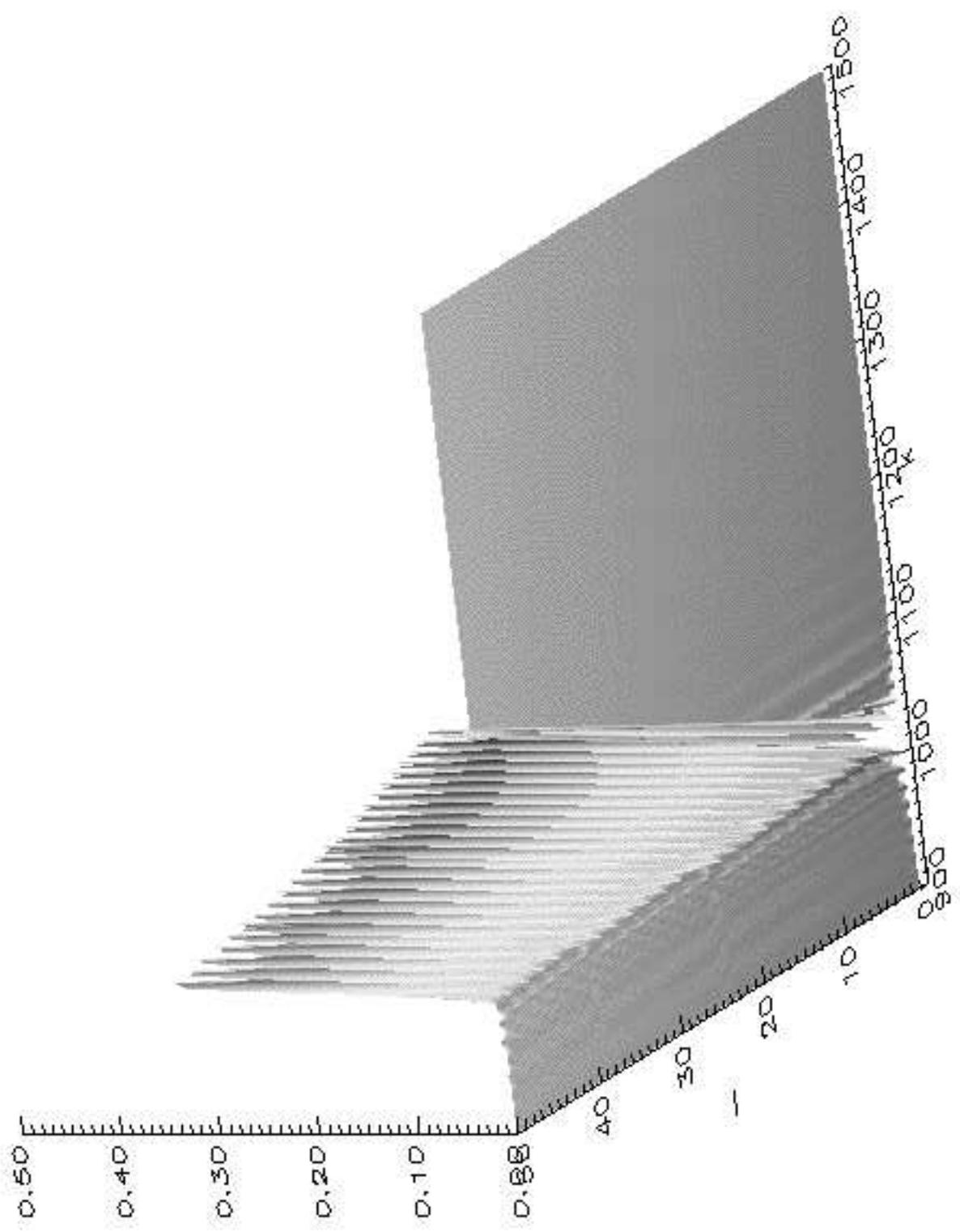
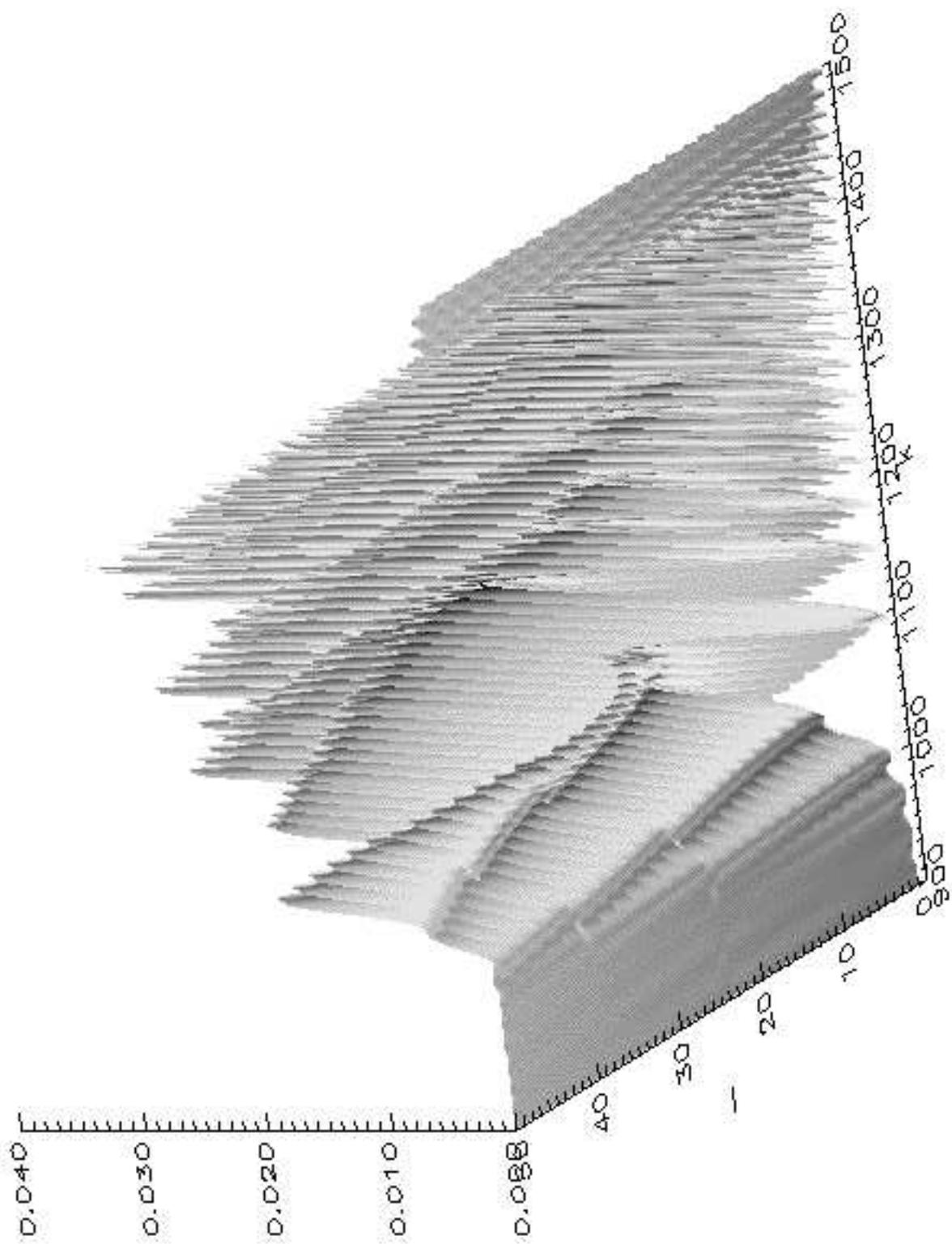
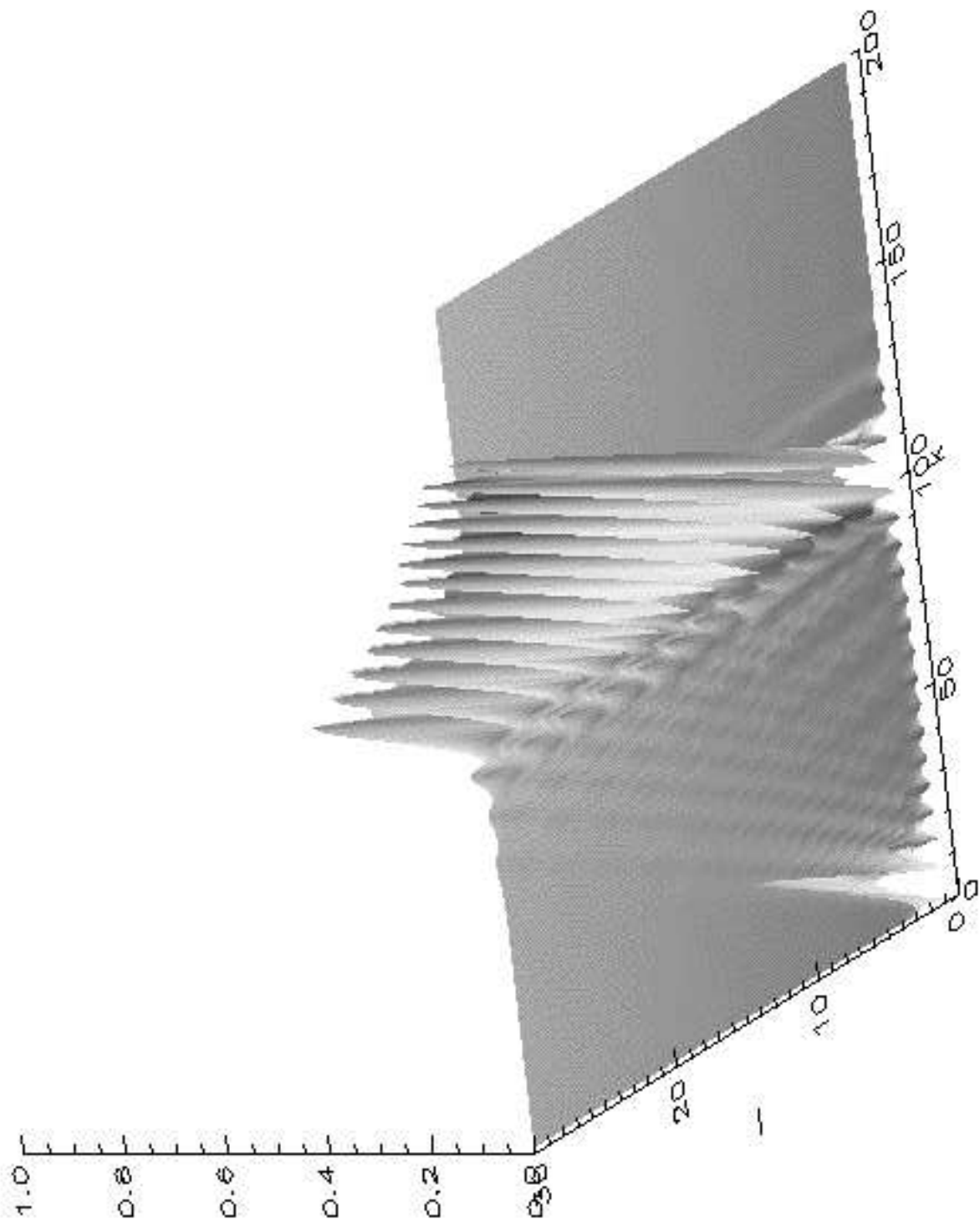
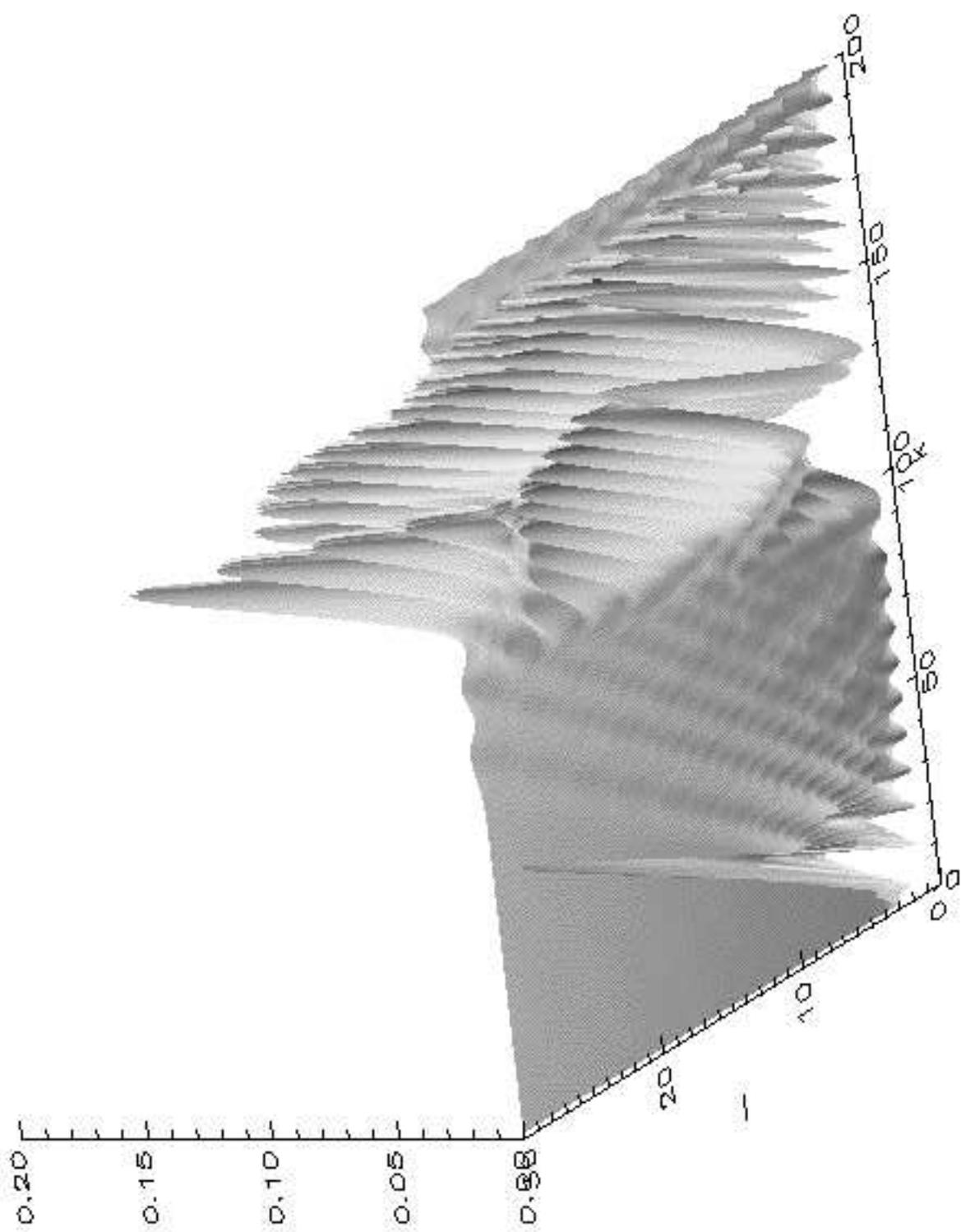
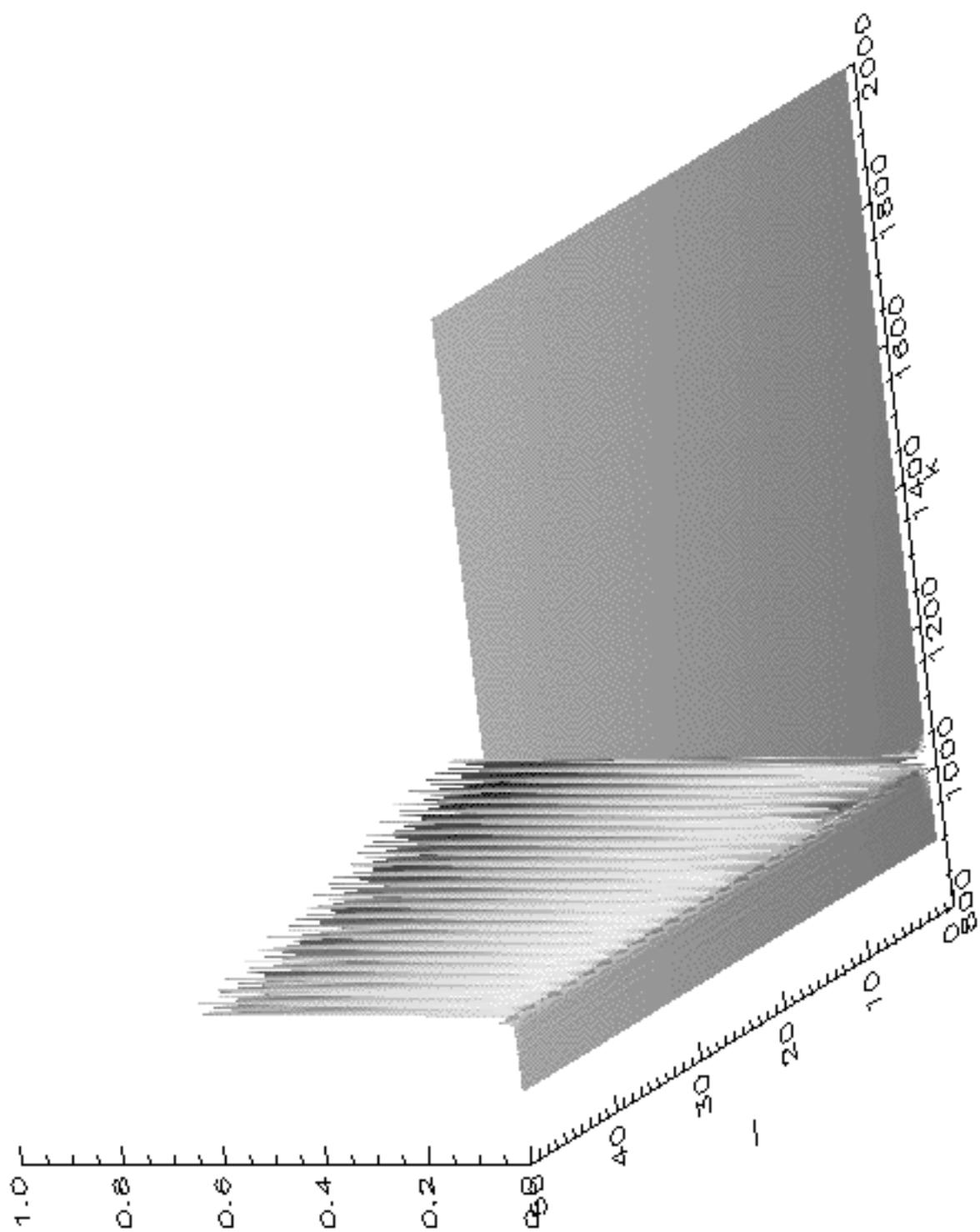
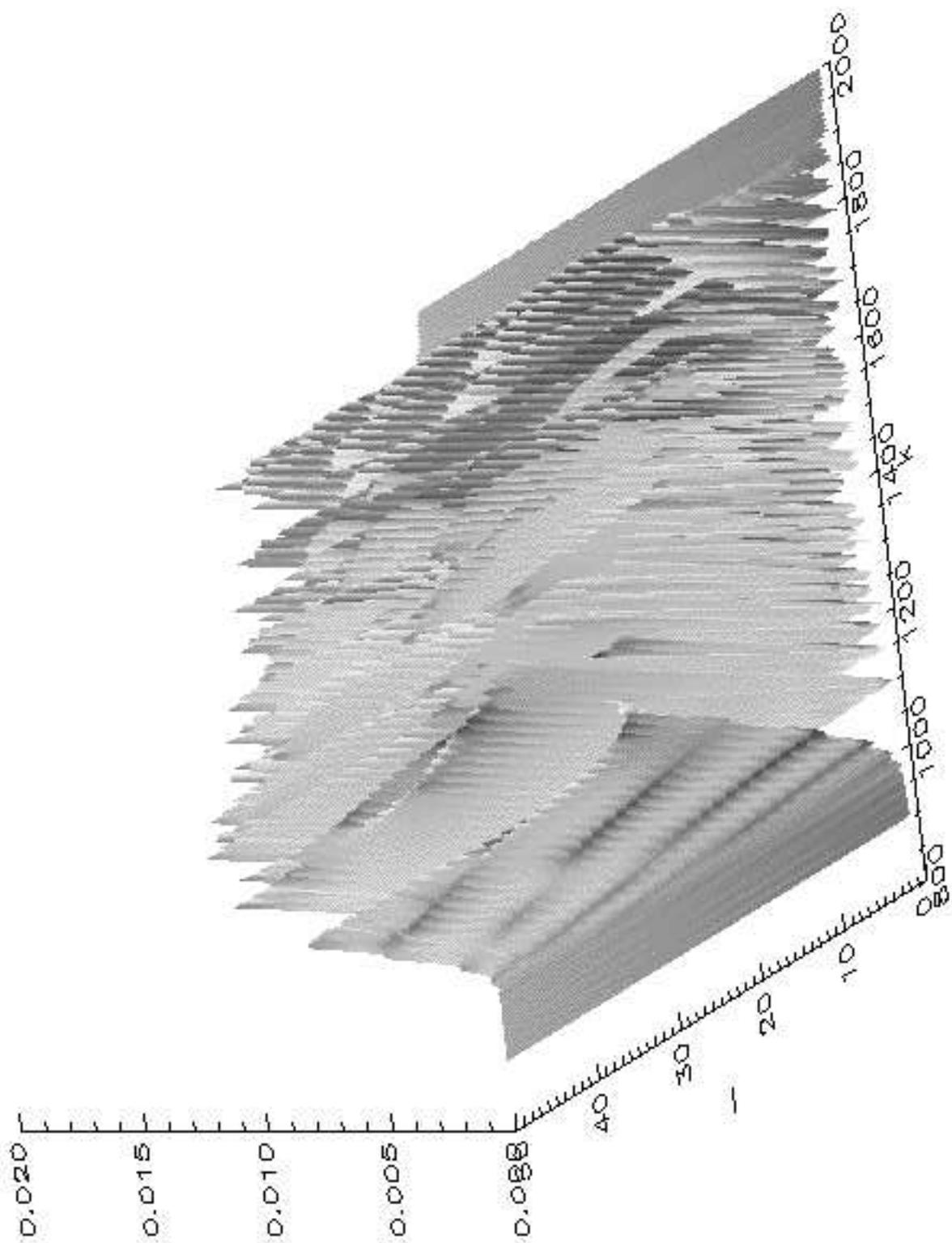
\end{document}